\documentclass[12pt,preprint]{aastex}
\usepackage{epstopdf, graphicx, natbib}
\bibpunct{(}{)}{,}{a}{}{;}

\def\lya{\ifmmode {\rm Ly}\alpha~ \else Ly$\alpha$~\fi}
\def\lyan{\ifmmode {\rm Ly}\alpha \else Ly$\alpha$\fi}
\def\lyb{\ifmmode {\rm Ly}\beta~ \else Ly$\beta$~\fi}
\def\lyg{\ifmmode {\rm Ly}\gamma~ \else Ly$\gamma$~\fi}

\def\caxiv{{{\rm Ca}\,{\sc xiv}~}}
\def\caxv{{{\rm Ca}\,{\sc xv}~}}

\def\civ{\ifmmode {\rm C}\,{\sc iv}~ \else C\,{\sc iv}~\fi}
\def\civn{\ifmmode {\rm C}\,{\sc iv}~ \else C\,{\sc iv}\fi}

\def\cvin{\ifmmode {\rm C}\,{\sc vi} \else C\,{\sc vi}\fi}

\def\nvn{N\,{\sc v}}

\def\nvii{N\,{\sc vii}~}

\def\siiv{{{\rm Si}\,{\sc iv}~}}

\def\sxiv{{{\rm S}\,{\sc xiv}~}}

\def\oi{{{\rm O}\,{\sc i}~}}
\def\oin{{{\rm O}\,{\sc i}}}
\def\oii{{{\rm O}\,{\sc ii}~}}

\def\oiv{{{\rm O}\,{\sc iv}~}}
\def\oin{{{\rm O}\,\hbox{{\sc i}}}}
\def\ovi{{{\rm O}\,{\sc vi}~}}
\def\ov{{{\rm O}\,{\sc v}~}}

\def\ovin{{{\rm O}\,{\sc vi}}}
\def\ovii{{{\rm O}\,{\sc vii}~}}
\def\oviii{{{\rm O}\,{\sc viii}~}}
\def\ovin{{{\rm O}\,{\sc vi}}}
\def\oviin{{{\rm O}\,{\sc vii}}}
\def\oviiin{{{\rm O}\,{\sc viii}}}
\def\neix{{{\rm Ne}\,{\sc ix}~}}
\def\neixn{{{\rm Ne}\,{\sc ix}}}
\def\nex{{{\rm Ne}\,{\sc x}~}}
\def\nexn{{{\rm Ne}\,{\sc x}}}

\def\fevii{{{\rm Fe}\,{\sc vii}~}}

\def\fevii-xii{{{\rm Fe}\,{\sc vii-xii}~}}

\def\fexviixxiin{{{\rm Fe}\,{\sc xvii-xxii}}}

\def\fexvii{{{\rm Fe}\,{\sc xvii}~}}
\def\fexviin{{{\rm Fe}\,{\sc xvii}}}
\def\fexviii{{{\rm Fe}\,{\sc xviii}~}}
\def\fexviiin{{{\rm Fe}\,{\sc xviii}}}

\def\fexxvi{{{\rm Fe}\,{\sc xxvi}~}}

\def\sixi{{{\rm Si}\,{\sc xi}~}}
\def\sixii{{{\rm Si}\,{\sc xii}~}}
\def\sixiin{{{\rm Si}\,{\sc xii}}}
\def\sixiii{{{\rm Si}\,{\sc xiii}~}}

\def\mgix{{{\rm Mg}\,{\sc ix}~}}
\def\mgxii{{{\rm Mg}\,{\sc xii}~}}
\def\mgxi{{{\rm Mg}\,{\sc xi}~}}
\def\mgxin{{{\rm Mg}\,{\sc xi}}}
\def\mgx{{{\rm Mg}\,{\sc x}~}}

\def\sixiii{{{\rm Si}\,{\sc xiii}~}}

\def\chandra{{\it Chandra}~}
\def\xmm{{\it XMM-Newton}~}

\def\ark564{\object{Ark~564}}
\def\mrk590{{\it Mrk~590}}
\def\ngc3783{{\it NGC~3783}}

\title{A Two-Phase Low-velocity Outflow in the Seyfert 1  
Galaxy \textbf{Ark~564 }}
\author{A. Gupta and S. Mathur\altaffilmark{1}}
\affil{Astronomy Department, Ohio State University, Columbus, OH 43210, USA}
\email{agupta@astronomy.ohio-state.edu}
\author{Y. Krongold}
\affil{Instituto de Astronomia, Universidad Nacional Autonoma 
de Mexico, Mexico City, (Mexico)}
\author{F. Nicastro}
\affil{Harvard-Smithsonian Center for Astrophysics, 
Cambridge, MA, 02138, USA}
\affil{Osservatorio Astronomico di Roma-INAF, 
Via di Frascati 33, 00040, Monte Porzio Catone, RM, (Italy)}

\altaffiltext{1}{Center for Cosmology and
Astro-Particle Physics, The Ohio State University, Columbus, OH 43210}

\begin{document}

\begin{abstract}
The Seyfert 1 galaxy \ark564 was observed with \chandra high energy
transmission gratings for 250 ks. We present the high resolution X-ray
spectrum that shows several associated absorption lines. The 
photoionization model requires two warm absorbers with 
two different ionization states ($logU=0.39\pm0.03$ and 
$logU=-0.99\pm0.13$), 
both with moderate outflow velocities ($\sim$100 km s$^{-1}$) and 
relatively low line of sight column densities ($logN_{H}=20.94$ 
and $20.11$ cm$^{-2}$). The high ionization phase 
produces absorption lines of \oviin, \oviiin, \neixn, \nexn,
\mgxin, \fexvii and \fexviii while the low ionization phase 
produces lines at lower energies (\ovi \& \oviin). 
The pressure--temperature equilibrium curve for the
\ark564 absorber does not have the typical ``S'' shape, even if the
metallicity is super-solar; as a result the two warm-absorber 
phases do not appear to be in pressure balance. This suggests that 
the continuum incident on the absorbing gas is perhaps different 
from the observed continuum.  We also estimated the mass outflow 
rate and the associated kinetic energy
and find it to be at most $0.009\%$ of the bolometric luminosity of
\ark564. Thus it is highly unlikely that these outflows provide
significant feedback required by the galaxy formation models.
\end{abstract}

\section{Introduction}

Outflows are ubiquitous in AGNs, manifested by high-ionization
absorption lines in X-rays and UV 
\citep[and references therein]{Reynolds1997, Crenshaw2003} 
and are perhaps related to the accretion process
\citep[e.g.,][]{Proga2007}. Understanding outflows is therefore as
important as understanding accretion itself. The X-ray absorbers,
commonly known as warm absorbers (WAs), have typical ionization parameter
$log~\xi$\footnote{The ionization parameter $\xi=L_{ion}/nr^{2}$, where
$L_{ion}$ is the ionizing luminosity between 1 Ryd and 1000 Ryd (1
Ryd=13.6 eV), {\bf n} is the number density of the material and {\bf r}
is the distance of the gas from the central source.} of $0-
2~$erg~s$^{-1}$, a column density of $N_{H}=10^{20}-10^{22}~$cm$^{-2}$ and
an outflow velocity of $100-1000~$km~s$^{-1}$, produced by warm ionized
gas ($T\sim10^{4}-10^{6}$ K; \citealt{Krongold2003}). The WAs are usually
detected in the  $0.3-2~$keV soft X-ray band by absorption lines of
Oxygen (\oviiin, \oviin, \ovin), Iron (\fevii-xii and \fexviixxiin), and
other highly ionized elements. Transitions by \civn, \nvn, and \ovi are
observed in both the X-ray and UV spectra of these sources with similar
outflow velocities, suggesting a connection between the narrow
absorption line systems in the UV and the X-rays WA \citep{Mathur1994,
Mathur1995, Kaspi2002, Krongold2003}.

Several phenomenological models have tried to explain AGN warm absorber
spectra showing multiple velocity components of multiple lines, and
after years of effort a consensus is growing.  In majority of the cases,
if not all, physical properties and kinematics of the absorber are well
determined and it can be described by at least two discrete ionization
components \citep{Detmers2011, Holczer2012}.  These components are
consistent with the same outflow velocity and appear to be in pressure
equilibrium, and so likely emerge from a multiphase wind
\citep[e.g.,][]{Krongold2003, Krongold2005, Krongold2007, Netzer2003,
Cardaci2009, Andrade2010}. The low-ionization phase (LIP) of the wind
produces UV and X-ray absorption lines, but the high-ionization phase
(HIP) is seen only in X-rays.

Despite ubiquitous detection and successful modeling of WA spectra
by multiple absorbing components, very little is still known about their
geometry and dynamical strength.  Where do these outflows originate?
Proposed locations span a wide range, of a factor of $10^{6}$ in radial
distance from the central ionizing source: the accretion disk \citep[as
suggested by the accretion disk wind models;][]{Proga2004}, the broad
line region \citep{Kraemer2005}, the obscuring torus
\citep{Dorodnitsyn2008, Krolik2001, Blustin2005} and to the narrow line
region \citep{Behar2003, Crenshaw2009}.  In principle, these outflows
could potentially provide a common form of AGN feedback required by
theoretical models of AGN-galaxy formation \citep[and references
therein]{Silk1998, Wyithe2003, Fabian2012}, although estimating their
mass outflow rate and the kinetic energy outflow rate depends critically
on the location of WAs. Some recent studies of WAs found that the
typical outflow velocity is a small fraction of the escape velocity and
that WAs do not carry sufficient mass/energy/momentum to be efficient
agents of feedback \citep{Mathur2009, Krongold2007, Krongold2010}.

With the goal of self consistent analysis and modeling of grating
spectra of WAs, we present here the results of our analysis of \chandra
archival data of \ark564.  The analysis of one \chandra observation of
this source done in 2000 (50 ks) has been published by
\citet{Matsumoto2004}; here we present new data of the three 2008
\chandra observations (250 ks total).

\section{Ark~564}
\ark564 is a bright, nearby, narrow-line Seyfert 1 (NLS1) galaxy, with
$z=0.024684$, $V=14.6$ mag (de Vaucouleurs et al. 1991), and
$L_{2-10~keV}=(2.4-2.8)\times10^{43}~$ergs~s$^{-1}$ \citep[and present
work]{Turner2001,Matsumoto2004}. It has been studied across all
wavebands \citep[e.g.,][]{Turner2001,Collier2001,
Shemmer2001,Romano2004} and shows large amplitude flux variations on
short time scales and a peculiar emission line-like feature near $1~$keV
\citep{Leighly1999, Turner2001, Comastri2001}.
\citeauthor{Matsumoto2004} analyzed the \chandra HETGS observation of
\ark564 (that carried out in 2000) with an exposure time of
$50.2~$ks. They modeled the hard X-ray spectrum with a power law of
photo-index of $2.56\pm0.06$ and fit the soft excess below $1.5~$keV
with a blackbody, of temperature $0.124\pm0.003~$keV.  They find some
evidence for a two phase WA with ionization parameters $log~\xi\sim1$
and $log~\xi\sim2$ and column density of $logN_{H}=21~$cm$^{-2}$.  They
find that the $1~$keV emission feature is not due to blends of several
narrow emission lines and suggest that it could be an artifact of the
warm absorber.  \citet{Brinkmann2007} studied the spectral variability
of the X-ray emission of the \ark564 using the $\sim100~$ks \xmm
observation and find that the ``power law plus bremsstrahlung'' model
describes the spectrum well at all times, with flux variations of both
components.

\citet{Papadakis2007} analyzed the \xmm EPIC data from \ark564 2005
observations. They found evidence for two phases of photoionized X-ray
absorbing gas with ionization parameter $log\xi\sim1$ and $log~\xi\sim2$
and column densities of $N_{H}\sim2$ and $5\times10^{20}~$cm$^{-2}$,
similar to the results of \citet{Matsumoto2004}.  They also detect an
absorption line at $\sim8.1~$keV in the low resolution CCD spectra and
assuming that this line corresponds to \fexxvi $K\alpha$, they suggest
the presence of highly ionized, absorbing material of
$N_{H}~>~10^{23}~$cm$^{-2}$ outflowing with relativistic velocity of
$\sim0.17c$.

Analyzing both the EPIC and RGS data from the same \xmm 2005 observations, 
\citet{Dewangan2007} find two warm absorber phases with 
ionization parameters $log~\xi\sim2$ and $log\xi<0.3$ and 
column densities of $N_{H}\sim4$ and $2\times10^{20}~$cm$^{-2}$ and outflow 
velocities of $300$ and $1000~$km~s$^{-1}$ respectively.

\citet{Smith2008} analyzed the combined \xmm RGS spectrum of \ark564
obtained from 2000 to 2005. They found three separate phases of
photoionized X-ray absorbing gas, with ionization parameters
$log~\xi=-0.86, 0.87, 2.56$ and column densities of $N_{H}=0.89, 2.41,
6.03\times10^{20}~$cm$^{-2}$ respectively, all with very low velocity
($-10\pm100~$km~s$^{-1}$). From the emission line analysis they found a
flow velocity of $-600~$km~s$^{-1}$ and claimed that the X-ray absorption
and emission originate in different regions.

\ark564 is also known to have a strong UV absorber, characterized
by \ovin, \sixiin, \siiv and \civ absorption lines 
\citep{Crenshaw2002,Romano2002}.

\section{Observations and Data Reduction}

\ark564 was observed with the \chandra High Energy Transmission Grating
Spectrometer (HETGS) on 2000 June for $50~$ks and between 2008 August 26
to 2008 September 6 for total duration of $250~$ks.  It was also
observed with Low Energy Transmission Grating Spectrometer (LETGS) on
2008 April 21 for 100 ks. Table 1 lists the observation log.
\citeauthor{Matsumoto2004} analyzed and presented the results of the $50~$ks
\chandra HETG observation done in 2000 June.  Here we report on the
\chandra archival 2008 observations of \ark564 made with both HETGS and
LETGS.

The HETGS consists of two grating assemblies, a high-energy grating
(HEG) and a medium-energy grating (MEG). The HEG bandpass is
$0.8-10~$keV and the MEG bandpass is $0.5-10~$keV but the effective area
of both instruments falls off rapidly at either end of the bandpass. We
performed the spectral analysis over $5-25~$\AA range. The LETG is combined
with the Advanced CCD Imaging Spectrometer-Spectroscopic (ACIS-S) array
or with the High Resolution Camera-Spectroscopic (HRC-S) array. The
\ark564 LETG observation was done with HRC-S array. The LETG/HRC-S has a
band pass of $0.07-7.3~$keV or $1.7-170$\AA, but due to the low S/N of
data on either ends, we restricted our spectral fitting to $5-40\AA$.

We reduced the data using the standard Chandra Interactive Analysis of
Observations (CIAO) software (v4.3) and Chandra Calibration Database
(CALDB, v4.4.2) and followed the standard Chandra data reduction
threads\footnote{\url{http://cxc.harvard.edu/ciao/threads/index.html}}.
For the Chandra ACIS/HETG observations, we co-added the negative and
positive first-order spectra and built the effective area files (ARFs)
for each observation using the \emph{fullgarf} CIAO script. Unlike
ACIS, the HRC does not have the energy resolution to sort individual
orders, and each spectrum contains contributions from all the
diffraction orders.  For the HRC/LETG observation we used the standard
ARF files of orders 1 through 6 and convolved them with the relevant
standard redistribution matrix files (RMF).

For HETG observations, we generated the light curves in $2~$ks bins for
the energy band ($0.3-10 ~$keV), as shown in figure 1. The time-average
count rate varies from $0.24~$cts~s$^{-1}$ to $0.32~$cts~s$^{-1}$ among
$2008$ observations. However, except for the initial $20~$ks of obsID
10575, the count rate of all observations are consistent with each
other, with average value of $0.31~$cts~s$^{-1}$.

We analyzed the spectra using the CIAO fitting package \emph{Sherpa}. As
noticed above, the HETG observations do not show large variations, so to
increase the signal to noise (S/N) of the spectrum we co-added the
spectra obtained with each observation and averaged the associated ARFs
using the ciao script \emph{add\_grating\_spectra}. This gave a total
net exposure time of $250~$ks for the MEG and HEG. We fit the MEG and
HEG data simultaneously, and discuss the LETG spectral analysis
separately in section 5. Throughout the paper we applied the $\chi^2$
minimization technique in the spectral analysis and the reported errors
are of $1\sigma$ significance for one interesting parameter.

\section{HETG Spectral Analysis}

\subsection{Continuum Modeling}

To model the intrinsic continuum of the source, we first fitted a simple
absorbed (Galactic $N_{H}=6.4\times10^{20}~$cm$^{-2}$;
\citet{Dickey1990}) power law with varying photon index and amplitude. A
single absorbed power law could not fit the data over the entire
range. We found an excess of flux in the spectrum at energies below
$\approx~1.5~$keV. \ark564 is known to have a strong soft continuum
\citep{Leighly1999,Turner1999}, so to fit this soft excess we added a
black-body component to the above mentioned simple power-law.  The fit
improved significantly ($\chi^{2}/d.o.f.=5882/3997$,
$\Delta\chi^{2}=515$) and figure 2 shows the continuum model fit to the
MEG spectrum.  We also plot the data:fit residuals, which show strong
absorption features consistent with the known WA of this source.

\subsection{Local $z=0$ Absorption}

The spectra of \ark564 show narrow absorption lines at zero
redshift. The \oi and \oii absorptions are attributed to the ISM
\citep{Wilms2000} and \oviin, \oviii and \fexvii absorption lines arise
in the circumgalactic medium (CGM) of our Galaxy or in the Local Group
\citep[and references therein]{Gupta2012}. We modeled all the
statistically significant local absorption features with Gaussian
components of fixed width of 1 m\AA; the line response function of the
grating is folded in the RMF file in each case.  The model of continuum
plus local absorption features (``Model A'') improves the fit by
$\Delta\chi^{2}=108$ for 10 fewer degrees of freedom. The measured
equivalent width (EW) and statistical significance of each line are
reported in Table 2 and labeled in figure 3.

\subsection{Intrinsic Absorption}

As reported above, the \ark564 spectrum shows many strong intrinsic
absorption features. We measured the position, EW and statistical
significance of all intrinsic absorption lines by fitting negative
Gaussians of fixed width of $1~$m\AA (Fig. 3, Table 2). All the \ark564
absorption features are blueshifted with respect to the source, implying
moderate outflow velocities of $82-239~$km~s$^{-1}$. We find a few lines
with no identification at $19.811\pm0.006~\AA$, $19.850\pm0.005~\AA$ and
$20.255\pm0.005~\AA$ in the observer frame. We marked these features
with green underlines in figure 3.

\subsection{Photoionization model fitting: PHASE}

We used the Photoionization model fitting code PHotoionized Absorption
Spectral Engine \citep[PHASE;][]{Krongold2003}, to model the warm
absorber features. The PHASE code self consistently models all the
absorption features observed in the X-ray spectra of AGNs.  At its
simplest, an absorption-line spectrum can be fit with PHASE using only
four input parameters: 1) the ionization parameter of the absorber {\bf
$U$}; 2) the equivalent hydrogen column density {\bf $N_{H}$}; 3) the
outflow velocity of the absorbing material {\bf $V_{out}$}; and 4) the
micro-turbulent velocity {\bf $V_{turb}$} of the material.  The
abundances have been set at the Solar values \citep{Grevesse1993}.  We
used the \ark564 spectral energy distribution (SED) from
\cite{Romano2004} to calculate the ionization balance of the absorbing
gas in PHASE. The SED constructed by Romano et al. is based on a
quasi-simultaneous multiwavelength campaign, and thus is the most
accurate overall SED for this source obtained to date. In the X-rays,
however, we use our own fits, as this radiation is the one responsible
for the ionization of the charge states producing absorption in the
X-ray band during our observations. We further connect the UV and the
X-ray data with a simple power law (a straight line in log-log space
connecting the last UV point and the first X-ray point).
 
The most prominent absorption lines in the HETGS \ark564 spectrum are
those of \mgxin, \nexn, \neixn, \fexviin, \fexviiin, \oviiin, \ovii and \ovi
at an outflow velocity of $\approx-100~$km~s$^{-1}$. We add a single
ionized absorbing component to Model A described above (we call it Model
B) to characterize this WA component.  This absorber has best fit
parameters of $logU=0.39\pm0.03$, $logN_{H}= 20.94\pm0.02$ cm$^{-2}$,
and an outflow velocity relative to systemic of $-94\pm13~$km~s$^{-1}$.
The fit gives a significant improvement over model A (Fig. 4;
$\chi^{2}/d.o.f.=4722/3982$, $\Delta\chi^{2}=1073$).  This absorber fits
the high ionization lines so we will refer to this absorber as the
``high-ionization phase (HIP)'' component.  The HIP component fits the
absorption features produced by ions such as \mgxin, \nexn, \neixn,
\fexviiin, \fexviin, \oviiin, and \ovii. \cite{Matsumoto2004} also report
an absorber of similar characteristics: ionization parameter
$log~U\sim1$ and absorption column $N_{H}=10^{21}~$cm$^{-2}$, derived
using the column densities of \oviiin, \neixn, \nex and \mgxin.

The single-absorber model does not fit the \ovi absorption line and it
also under-predicts the \ovii absorption (Fig. 5, red curve).  This
suggests the presence of another absorber with lower ionization
state. To fit the residual features we added another absorber to our
previous model defining Model C.  An absorber with ionization parameter
$logU=-0.99\pm0.13$, $logN_{H}= 20.11\pm0.06$ cm$^{-2}$, and an outflow
velocity of $\sim-137\pm37~$km~s$^{-1}$ successfully fits the low
ionization lines, including \ovi and \ovii (Fig. 5 \& Fig. 6).  We call
this component ``lower-ionization phase (LIP)'' absorber.  This model
gives a $\chi^{2}/d.o.f.=4674/3979$; $\Delta\chi^{2}=48$, significantly
improving over the single-absorber model B. An F-test gives a higher
than $99.999$\% confidence for the presence of this absorber.  The ionic
column densities predicted by our best fit two-ionized absorber model
(Model C) are listed in Table 4.  As can be inferred from Table 4,
except for \oviin, the LIP and HIP components predict absorption from
different ions but at similar velocities. This suggest that the \ark564
absorbers may be present in the same outflowing multi-phase medium. We
come back to this in section 6.3.

To extrapolate the two component ionized absorber model of the MEG
spectrum of \ark564 to lower wavelengths, we simultaneously fit the HEG
and MEG spectra (Table 3, Fig. 7). The HEG intrinsic absorption features
are well fitted with HIP absorber, including \sixiii and \mgix lines
which were not detected in the MEG spectrum due to low signal to
noise. The best-fit model parameters of the MEG$+$HEG fit are consistent
with the MEG-only fit.

Though most of the \ark564 intrinsic absorption features are well fitted
with two warm absorbers, the model does not fit the unidentified
absorption lines mentioned in section 4.3.  The identification,
evaluation and interpretation of these features is be discussed in
detail in a companion paper (Gupta et al. 2013b).

\section{LETGS Spectral Analysis}

\ark564 was also observed with LETGS in April 2008. \citet{Ramirez2013}
analyzed that observation and found a very weak absorption line of \ovii
K$\alpha$ but a strong feature at $18.62~\AA$ (at the wavelength of
\ovii K$\beta$). For this reason they identified the absorption feature
at $18.62~\AA$ as blueshifted \oviii K$\alpha$ with velocity
$\sim5500~$km~s$^{-1}$. They further modeled the $17-25~\AA$ spectral
region with two high ionization absorbing components with
$log(\xi)\sim3$, one at $v\sim0~$km~s$^{-1}$ and one at
$v\sim5500~$km~s$^{-1}$.  Since we found no evidence of an outflow with
velocity $\sim5500~$km~s$^{-1}$ in the $2008$ HETGS spectra, we
reanalyzed the LETGS observation to check for the consistency with our
model derived from the HETG data.

To fit the \ark564 LETGS spectrum continuum, we used the same model as
for HETG data (a power law plus black-body). The best fit power law
photon-index and temperature of black-body are reported in Table 3.  We
observed that the flux ($2-10~$keV) of the source varied from
$2.48\pm0.11 \times 10^{-11}~$erg~s$^{-1}~$cm$^{-2}$ to $2.79\pm0.15
\times 10^{-11}~$erg~s$^{-1}~$cm$^{-2}$ between the HETG and LETG
observations. To fit the intrinsic absorption features, we start with
the two component ionized absorber model obtained for HETG spectra. This
model fits the absorption features of \fexviiin, \fexviin, \nexn, \neix
reasonably well, but overestimates the \ovii k$\alpha$ line strength
($\chi^{2}/do.f.=1426/1190$; Fig. 8 \& Fig. 9). Allowing the PHASE
parameters to vary freely, the fit improved considerably
($\chi^{2}/do.f.=1370/1184$; Fig. 10). The best fit parameters of LETGS
WA model are reported in Table 3. This model fits the narrow absorption
due to \ovii $K\alpha$, but leaves the residuals at $19.1\AA$,
corresponding to \ovii $K\beta$ (Fig. 9). We also tried to fit the LETG
spectrum with models suggested by \cite{Ramirez2010}, but the fit was
not good ($\chi^{2}/do.f.=1936/1190$). Though this model well fits the
line at 19.1~\AA (observed frame), it also predicts other absorption
lines which are inconsistent with the data. As noted by
\cite{Ramirez2010}, the absorption line at $19.1\AA$ is too strong to be
by \ovii $K\beta$ and could be, in part, due to a transient high
velocity outflow component (Gupta et al. 2013b, companion paper).

Between the LETG and HETG warm absorber models, the HIP component 
parameters are consistent within errors. However, the LETG LIP has 
lower ionization parameter and higher column in comparison to HETG LIP. 
We also note that the column densities of highly ionized ions 
(\oviiin, \fexviin, \fexviiin, \neixn, \nexn, \mgxin, \mgxii and \sixiin) 
are higher while for those of less ionized ions (\ovii and \ovin) 
are smaller in the HETG observation than in LETG (Table 4). 
Both the instruments have good response in the spectral region 
where these lines are detected, so the observed differences cannot 
be due to instrumental artifacts; what is observed is the real 
variability in the WA properties between the two observations from 
2008 April (LETG) to August/September (HETG).

\section{Discussion}

\subsection{The Connection between UV and X-ray Absorbers}

\ark564 was observed with HST (STIS) and FUSE during May-July 2000 and
June 2001 respectively. \citet{Crenshaw2002} and \citet{Romano2002}
detected intrinsic absorption lines blueshifted by
$\approx-190~$km~s$^{-1}$ and $\approx-120~$km~s$^{-1}$ respectively,
similar to that of the X-ray WAs. \citeauthor{Crenshaw2002} modeled the
UV data with a single absorber of ionization parameters $logU$ of
$\sim0.033$ and column density $logN_{H}$ of $\sim21.2$ cm$^{-2}$.  The
UV absorption model also predicted the column densities of \ovii and
\oviii of $<2.2 \times 10^{17}~$cm$^{-2}$ and $<1.1 \times
10^{16}~$cm$^{-2}$ respectively \citep{Romano2002}. The \ovii column
density measured from the X-ray WA models $(0.3-1.3)\times
10^{17}~$cm$^{-2}$ is consistent with UV upper limits. The total
HIP$+$LIP \oviii column density of $(2.0-2.9) \times 10^{17}~$cm$^{-2}$
is an order of magnitude higher than the UV estimates. However, the LIP
\oviii column density $=(3.0-4.6) \times 10^{15}~$cm$^{-2}$ is in
agreement with UV models. The consistency between the X-ray LIP absorber
outflow velocity, hydrogen column density, and ionic column densities
with the UV absorber model suggests that both are the same absorber. The
HIP absorber, on the other hand, is different from the UV absorber, as
expected.

\citeauthor{Romano2002} from FUSE observations of \ark564 measured the
intrinsic \ovi column density of $(5.70-6.01) \times 10^{15}~$cm$^{-2}$.
The upper limit on \ovi column density measured from UV data is much
smaller than our lower limit on \ovi column of $1.0 \times
10^{16}~$cm$^{-2}$ in X-rays. This could be due to the variability of
the WA between the two observations. We note, however, that X-ray and UV
\ovi column densities have been found to be discrepant in other AGN
absorption systems \citep[e.g.,][]{Krongold2003} and in redshift zero
absorption \citep{Williams2006}.

\subsection{Estimates on mass and energy outflows rates}

Using the values of warm-absorber parameters such as column density,
ionization parameter and outflowing velocities, we can give a rough
estimate of the mass outflow rate ($\dot{M}_{out}$) and the kinetic
energy carried away by the warm absorbing winds (i.e. kinetic
luminosity, $L_{K}$). But before we can measure the mass and energy
outflow rates, we must know the location of the absorber.  However, in
the equation for the photoionization parameter ($U\propto L/ n_e R^2$),
the radius of the absorbing region (R) is degenerate with the density
($n_{e}$). In principle we can put constraints on the distance $R$ by
measuring the density of the WA with variability analysis 
\citep[e.g.,][]{Krongold2007}.  Using this technique, 
\citeauthor{Krongold2007} managed to determine the absorber 
density in NGC4051 and so the distance.  For \ark564,  
 no such study is available in literature and in the present work 
the source does not show any significant variability either. 
Therefore, we will only set limits on the mass and energy outflow 
rates using the expression derived in \citeauthor{Krongold2007}, 
$\dot{M}_{out} \approx 1.2 \pi m_{p} N_{H} v_{out}r$.

The estimate of maximum distance from the central source can be derived
assuming that the depth $\Delta r$ of the absorber is much smaller than
the radial distance of the absorber ($\Delta r << r$) and using the
definition of ionization parameter ($U=\frac{Q(H)}{4\pi r^{2}n_{H}c}$),
i.e.  $r \leq r_{max}=\frac{Q(H)}{4 \pi U N_{H} c}$. In several papers
lower limit on the absorber distance was determined assuming that the
observed outflow velocity is larger than the escape velocity at $r$:
i.e. $r \geq r_{min}=\frac{2GM_{BH}}{v_{out}^{2}}$. However, as shown in
\citet{Mathur2009} WA outflow velocities are usually lower than the
escape velocities, so cannot really be used to derive a lower limit on
$r$. Using the best fit values of ionization parameter and column density, 
we estimated the upper limits on HIP and LIP absorber locations of 
$r_{HIP} < 40~$pc and $r_{LIP} < 6~$kpc respectively, which are not
very interesting limits. Using the above equation and outflow 
velocities of 94~km~s$^{-1}$ and 137~km~s$^{-1}$, we obtain the mass
outflow rates of $\dot{M}_{out} < 6.4 \times 10^{24}~$g~s$^{-1}$ and
$\dot{M}_{out} < 2.2 \times 10^{26}~$g~s$^{-1}$ for HIP and LIP absorbers 
respectively.  Similarly we obtained the constraints on kinetic
luminosity of the outflows of $\dot{E}_{K} < 2.8 \times
10^{38}~$erg~s$^{-1}$ and $\dot{E}_{K} < 2.1 \times 10^{40}~$erg~s$^{-1}$
for the HIP and LIP absorbers respectively.

In comparison to the \ark564 bolometric luminosity of $2.4 \times
10^{44}~$erg~s$^{-1}$ \citep{Romano2002}, the total kinetic luminosity 
of these outflows is $\dot{E}_{K}/L_{bol} < 0.0001$\% for the 
HIP and $<0.009\%$ for the LIP. Thus it is very unlikely that these 
outflows significantly affect
the local environment of the host galaxy. The AGN feedback models
typically required $0.5-5$\% of the bolometric luminosity of an AGN to
be converted into kinetic luminosity to have a significant 
impact on the surrounding environment \citep{Hopkins2010, Silk1998, SO04}.

\subsection{Pressure Balance between LIP and HIP}

The presence of two different absorbing components with different
temperatures but similar outflow velocity suggests that the absorber
could arise from two phases of the same medium
\citep[e.g.,][]{Elvis2000, Krongold2003}. This is further supported by
the fact that multiple components of the ionized absorber are found in
pressure balance \citep[e.g.,][]{Krongold2003, Krongold2005,
Krongold2007, Cardaci2009, Andrade2010, Zhang2010}.  This result has
proven valid against different methodologies and codes used in the
analysis \citep{Krongold2013}.  To investigate whether or not the
absorbing components of Ark 564 are in pressure balance, we generated
the pressure-temperature equilibrium curve (also known as the
``S-curve'' \citep{Krolik1981}, for the SED used in our analysis: Fig. 
11). Interestingly, we find that the equilibrium curve of \ark564 does
not have the typical ``S'' shape where multiple phases can exist in
pressure equilibrium, because there are no regions of instability (for
all points in the plane the derivative of the curve is positive).

Accepted at face value, this result implies that the absorber in \ark564
is not in pressure balance and thus is not forming a multiphase medium,
a result at odds with previous evidence on warm absorbers. While we
cannot rule out this possibility, there are several arguments pointing
towards a multiphase medium.  We note that the two different absorbing
components in \ark564 have the same kinematics, which suggests that they
are related. If they share the location (the most reasonable
assumption), there must be a gradient of pressure between them, given
that the LIP pressure is over 5 times larger than that of the
HIP. Therefore, these two components should move on the S-curve to form
a single component in a time comparable to the free expansion time,
given by $t_{exp}=\Delta R/V_{s}$ (where $\Delta R$ is the thickness of the
absorber and $V_{s}$ the speed of sound in the medium). The flow time of
the components (i.e. the time in which the components cross our line of
sight to the source) is given to first order as $t_{flow}=R/V_{out}$
(where $R$ is the distance from the illuminated face of the absorbing
material to the ionizing source and $V_{out}$ its outflow velocity). For
the warm absorber in \ark564 $V_{s}\sim V_{out}$ (specially for the HIP
that is hotter). It follows that $t_{exp}/t_{flow}=\Delta R/R < 1$. 
Then, the free expansion time is smaller than the flow time, and the
two phase should dissolve into a single component before moving out from
our line of sight, which is clearly not consistent with the data.  Thus,
if the two phases are not in pressure balance they should not be connected,
and the similar kinematics in this, and in many other sources, would
have to be considered a coincidence.

Alternatively, the sources might be in pressure balance forming a
multiphase medium, but there might be additional heating and/or cooling
processes acting on the gas, changing the shape of the S-curve, but not
the ionization balance. The most obvious parameter for this is the gas
metallicity. \citet{Komossa2001} showed that the shape of the
equilibrium curve not only depends upon the SED of the source, but also
on the metallicity of the absorber, which affects the cooling of the
gas. They further show that super-solar abundances restore the
equilibrium zone in steep spectrum sources and increase the pressure
range where a multiphase equilibrium is possible.  \citet{Fields2007} 
showed that this is indeed the case for the ionized absorber in Mkn 279.

Since \ark564 also has a steep spectrum and a monotonically rising ``S''
curve, we generated a new pressure-temperature curve with super-solar
metallicity of 10 solar, shown as a dashed curve in figure 11. This is a
good assumption as supersolar metallicity has been suggested for this
source \citep{Romano2004}. Even with super-solar metallicity, the
``S-curve'' is very steep, without regions where multiple components can
coexist in pressure equilibrium. Thus, even having high metallicity does 
not solve the problem of finding the warm absorber components out of 
pressure balance and below we speculate on possible reasons. 

The continuum X-ray spectrum of \ark564 is not only steep 
($\Gamma > 2.4$), it also
has an additional prominent soft excess, similar to the behavior seen in
sources with steep soft X-ray spectra (e.g., NGC 4051,
\citealt{Komossa2001}; Mrk279, \citealt{Fields2007}). In fact,
additional modeling shows that the main reason for a steep ``S-curve''
is the extra heating produced by the soft excess (particularly in the
LIP). If the soft excess continuum is not impinging directly on the
absorbing gas, perhaps because it is the result of reflection toward our
line of sight, then the two components would be in pressure
balance. Other possibilities to produce a multi-valued S-curve, and
LIP/HIP components in pressure balance include a weaker IR radiation
field illuminating the material than the one observed (Krolik and Kriss
2001) or additional (more exotic) sources of heating at high
temperatures, such as those discussed in \citet{Krolik1981}. Therefore,
if warm absorbers are indeed a multiphase medium in pressure
equilibrium, it is likely that the overall radiation field impinging on
the gas is different than the one observed. This effect might be
stronger in sources with steep soft X-ray spectra. We note, however,
that this suggestion is speculative; we cannot prove it or rule it out.
We also note that photoionization models of the Broad Line Region demand
that the ionizing continuum is different than the one observed
\citep[and references therein]{Binette2008}. If warm absorbers are
indeed in pressure balance, their S-curves can be used for a better
understanding of the physical properties and the processes acting on the
material \citep{Komossa2001, Chakravorty2012, Krongold2013}.
 
\section{Summary}

Our best fit model of intrinsic absorption of NLS1 galaxy \ark564
requires a two-phase warm absorber with two different ionization states
(HIP and LIP). Both the absorbers are outflowing at low velocities of
order of $\sim100~$km~s$^{-1}$.  The HIP absorber reproduces  
most of the spectral features observed in the HETG spectra (\oviiin, \neixn,
\nexn, \mgxin, \fexvii and \fexviiin) except for a few at lower energies
(\ovii and \ovin), which are modeled by the LIP component. The
pressure--temperature equilibrium curve for the \ark564 warm absorber
does not have the typical ``S'' shape, even if the metallicity is
super-solar; as a result the two WA phases do not appear to be in
pressure balance. We speculate that the continuum incident on the
absorbing gas is perhaps different from the observed continuum. We
observe clear variability in the WA properties between the 2008 HETG
observations and previous observations which could be in response to
the change in continuum or the absorbing clouds passing our sight-line;
the large time gap between observations does not allow us to distinguish
between the two possibilities.

We also estimated the mass outflow rate and associated kinetic energy
assuming a biconical wind model \citep{Krongold2007} and find that it
represents a tiny fraction of the bolometric luminosity of \ark564. Thus
it is highly unlikely that these outflows provide significant feedback
required by the galaxy formation models.

\noindent
{\bf Acknowledgement:} Support for this work was provided by the
National Aeronautics and Space Administration through Chandra Award
Number TM9-0010X issued by the Chandra X-ray Observatory Center, which
is operated by the Smithsonian Astrophysical Observatory for and on
behalf of the National Aeronautics Space Administration under contract
NAS8-03060. YK acknowledges support from CONACyT 168519 grant and 
UNAM-DGAPA PAPIIT IN103712 grant.

\bibliography{apj}

\begin{thebibliography}{}
\bibitem[Andrade-Velazquez et al.(2010)]{Andrade2010}Andrade-Vel\'{a}zquez, 
  M. et al. 2010, \apj, 711, 888


\bibitem[Behar et al.(2003)]{Behar2003}Behar, E. et al. 2003, \apj, 598, 232

\bibitem[Blustin et al.(2005)]{Blustin2005}Blustin, A. J., Page, M. J., 
Fuerst, S. V., Branduardi-Raymont, G., \& Ashton, C. E. 2005, \aap, 431, 111

\bibitem[Binette \& Krongold(2008)]{Binette2008}Binette, L. \& Krongold, Y. 
2008, \aap, 477, 413

\bibitem[Brinkmann et al.(2007)]{Brinkmann2007}Brinkmann, W., 
Papadakis, I. E., \& Raeth, C. 2007, \aap, 465, 107B

\bibitem[Cardaci et al.(2009)]{Cardaci2009}Cardaci, M. V. et al. 2009, \aap, 
505, 541

\bibitem[Chakravorty et al.(2012)]{Chakravorty2012}Chakravorty, S., 
 Misra, R., Elvis, M., Kembhavi, A.K. \& 
Ferland, G. et al. 2010, \mnras, 422, 637 

\bibitem[Collier et al.(2001)]{Collier2001}Collier, S. et al. 
2001, \apj, 561, 146

\bibitem[Comastri et al.(2001)]{Comastri2001}Comastri, A. 
et al. 2001, \aap, 365, 400

\bibitem[Costantini et al.(2007)]{Costantini2007}Costantini, 
E., Gallo, L. C., Brandt, W. N., Fabian, A. C., \& Boller, T. 2007, 
\mnras, 378, 873

\bibitem[Crenshaw et al.(1999)]{Crenshaw1999}Crenshaw, D. M., Kraemer, S. B., 
Boggess, A., Maran, S. P., Mushotzky, R. F., \& Wu, C.-C. 1999, \apj, 516, 750

\bibitem[Crenshaw et al.(2002)]{Crenshaw2002}Crenshaw, D. M., Kraemer, S. B., 
Turner, T. J., et al. 2002, \apj, 566, 187

\bibitem[Crenshaw et al.(2003)]{Crenshaw2003}Crenshaw, D. M., Kraemer, S. 
B., \& George, I. M. 2003, A\&A Rev., 41, 117

\bibitem[Crenshaw et al.(2009)]{Crenshaw2009}Crenshaw, D. M. et al. 
2009, \apj, 698, 281

\bibitem[Detmers et al.(2008)]{Detmers2008}Detmers, R. G. et al. 2008, 
\aap, 488, 67

\bibitem[Detmers et al.(2011)]{Detmers2011}Detmers, R.G. et al. 
2011, \aap, 534, 38

\bibitem[Dewangan et al.(2007)]{Dewangan2007}Dewangan, G. C., 
Griffiths, R. E., Dasgupta, S., \& Rao, A. R. 2007, \apj, 671, 1284

\bibitem[Dickey et al.(1990)]{Dickey1990}Dickey, J.M. 1990, ASSL, 161,473

\bibitem[Dorodnitsyn et al.(2008)]{Dorodnitsyn2008}Dorodnitsyn, A., 
Kallman, T., \& Proga, D. 2008, \apj, 687, 97

\bibitem[Elvis et al.(2000)]{Elvis2000} Elvis, M. 2000, \apj, 545, 63

\bibitem[Fabian (2012)]{Fabian2012}Fabian, A.C., arXiv:1204.4114v1

\bibitem[Fields et al.(2007)]{Fields2007}Fields, D.L., Mathur, S., 
Krongold, Y., Williams, R. \& Nicastro, F. 2007, \apj, 666, 828

\bibitem[Garcia et al.(2005)]{Garcia2005}Garcia, J. et al. 2005, \apjs, 158, 68

\bibitem[Grevesse et al.(1993)]{Grevesse1993}Grevesse, N., Noels, A., \& 
Sauval, A. J. 1993, \aap, 271, 587

\bibitem[Gupta et al.(2012)]{Gupta2012}Gupta, A., Mathur, S., 
Krongold, Y., Nicastro, F., \& Galeazzi, M. 2012, \apj, 756, L8


\bibitem[Holczer \& Behar(2012)]{Holczer2012}Holczer,T. \& 
Behar, E. 2012, \apj, 747, 71

\bibitem[Hopkins \& Elvis(2010)]{Hopkins2010}Hopkins, P. \& Elvis, M. 
MNRAS, 401, 7

\bibitem[Kaspi et al.(2002)]{Kaspi2002}Kaspi, S., et al. 
2002, \apj, 574, 643


\bibitem[Komossa \& Mathur(2001)]{Komossa2001}Komossa, S \& 
Mathur, S. 2001, \aap, 374, 914

\bibitem[Kraemer et al.(2002)]{Kraemer2002}Kraemer, S. B. et al. 2002, 
\apj, 577, 98

\bibitem[Kraemer et al.(2005)]{Kraemer2005}Kraemer, S. B., et al. 2005, 
\apj, 633, 693

\bibitem[Krolik et al.(1981)]{Krolik1981}Krolik, J. H., McKee, C. F., 
Tarter, C. B. 1981, \apj, 249, 422

\bibitem[Krolik \& Kriss(2001)]{Krolik2001}Krolik, J. H., \& 
Kriss, G. A. 2001, \apj, 561, 684

\bibitem[Krongold et al.(2003)]{Krongold2003}Krongold, Y., 
Nicastro, F., Brickhouse, N.S., Elvis, M., Liedahl D.A. \& Mathur, S. 2003,
\apj, 597, 832 

\bibitem[Krongold et al.(2005)]{Krongold2005}Krongold, Y., 
Nicastro, F., Elvis, M.,  Brickhouse, N. S., Mathur, S., \& Zezas, A. 2005, 
\apj, 620, 165

\bibitem[Krongold et al.(2007)]{Krongold2007}Krongold, Y. et al. 2007, 
\apj, 659, 1022

\bibitem[Krongold et al.(2010)]{Krongold2010}Krongold, Y., Binette, L., \& 
HernNandez-Ibarra, F. 2010, \apj, 724L, 203K
\bibitem[Krongold et al.(2013)]{Krongold2013}Krongold, y. et al. 
2012, in preparation

\bibitem[Leighly et al.(1999)]{Leighly1999}Leighly, K.M. 1999, \apjs, 125, 317

\bibitem[Mathur et al.(1994)]{Mathur1994}Mathur, S., Wilkes, B., 
Elvis, M., \& Fiore, F. 1994, ApJ, 434, 493

\bibitem[Mathur et al.(1995)]{Mathur1995}Mathur, S., Elvis, M., \& 
Wilkes, B. 1995, \apj, 452, 230

\bibitem[Mathur et al.(2003)]{Mathur2003}Mathur, S., Weinberg, D. H., \& 
Chen, X. 2003, \apj, 582, 82

\bibitem[Mathur et al.(2009)]{Mathur2009}Mathur, S., Stoll, R., Krongold, Y., 
Nicastro, F., Brickhouse, N., \& Elvis, M. 2009, AIPC, 1201, 33

\bibitem[Matsumoto et al.(2004)]{Matsumoto2004}Matsumoto, C., 
Leighly, K. M., \& Marshall, H. L. 2004, \apj, 603, 456

\bibitem[McKernan et al.(2007)]{McKernan2007}McKernan, B., Yaqoob, T., \& 
Reynolds, C. S. 2007, \mnras, 379, 1359

\bibitem[Netzer et al.(2003)]{Netzer2003}Netzer, H. et al. 2003, 
  \apj, 599, 933

\bibitem[Papadakis et al.(2007)]{Papadakis2007}Papadakis, I. E., 
Brinkmann, W., Page, M. J., McHardy, I., \& Uttley, P. 2007, \aap, 461, 931

\bibitem[Proga \& Kallman(2004)]{Proga2004}Proga, D., \& 
Kallman, T. R. 2004, \apj, 616, 688

\bibitem[Proga(2007)]{Proga2007}Proga, D. 2007, ASPC, 373, 267


\bibitem[Ramirez(2013)]{Ramirez2013}Ramirez, J. 2013, A\&A, 551, A95

\bibitem[Reynolds(1997)]{Reynolds1997}Reynolds, C. S. 1997, MNRAS, 286, 513

\bibitem[Romano et al.(2002)]{Romano2002}Romano, P., 
Mathur, S., Pogge, R. W., 
Peterson, B. M., \& Kuraszkiewicz, J. 2002, \apj, 578, 64

\bibitem[Romano et al.(2004)]{Romano2004}Romano, P., Mathur, S., \& 
Turner, T. J. 2004, \apj, 602, 635

\bibitem[Scannapieco \& Oh(2004)]{SO04}
Scannapieco, E. \& Oh, S., 2004 ApJ, 608, 62

\bibitem[Shemmer et al.(2001)]{Shemmer2001}Shemmer, O. 
et al. 2001, \apj, 561, 162

\bibitem[Silk \& Rees(1998)]{Silk1998}Silk, J. \& Rees, 
M.J. 1998, \aap, 331, 1
	
\bibitem[Smith et al.(2008)]{Smith2008}Smith, R. A. N., Page, M. J., \& 
Branduardi-Raymont, G. 2008, \aap, 490, 103

\bibitem[Turner et al.(1999)]{Turner1999}Turner, T. J., George, I. M., 
Nandra, K., \& Turcan, D. 1999, \apj, 524, 667

\bibitem[Turner et al.(2001)]{Turner2001}Turner, T. J., Romano, P., 
George, I. M., et al. 2001, \apj, 561, 131

\bibitem[Williams et al.(2006)]{Williams2006}Williams, R.J., 
Mathur, Smita, Nicastro, F., \& Elvis, M. 2006, \apj, 642, 95

\bibitem[Wilms et al.(2000)]{Wilms2000}Wilms, J., Allen, A., \& 
McCray, R. 2000, \apj, 542, 914

\bibitem[Wyithe \& Loeb(2003)]{Wyithe2003}Wyithe, J.S.B., \& 
Loeb, A. 2003, \apj, 595, 614

\bibitem[Yao et al.(2009)]{Yao2009}Yao, Y. et al. 2009, \apj, 696, 1418

\bibitem[Zhang et al.(2010)]{Zhang2010}

\end{thebibliography}

\clearpage

\begin{deluxetable}{ccc}
\tabletypesize{\scriptsize}
\tablecaption{Ark564 \chandra Observation Log.}
\tablewidth{0pt}
\tablehead{
\colhead{ID} & \colhead{Start Time} & \colhead{Exposure $(sec)$}
}
\startdata
& HETGS-ACIS-S & \\
\tableline
9899 & 2008-08-28 12:51:50 & 84077\\
9898 & 2008-09-06 02:14:14 & 99528\\
10575 & 2008-09-07 22:05:20 & 62216\\
\tableline
& LETGS-HRC-S & \\
\tableline
9151 & 2008-04-24T06:04:06 & 99962\\
\enddata
\end{deluxetable}

\clearpage

\begin{deluxetable}{lccccc}
\tabletypesize{\scriptsize} \tablecaption{Absorption lines observed in
the Ark~564 \chandra HETG-MEG spectra.}  
\tablewidth{0pt} \tablehead{
\colhead{$\lambda_{obs}$} & \colhead{EW} & \colhead{$\Delta
\chi^{2}~$\tablenotemark{a}} & \colhead{Ion Name}\tablenotemark{b} &
\colhead{$\lambda_{rest}$\tablenotemark{c}} & 
\colhead{$v_{out}$}\\ \colhead{$\AA$} &
\colhead{$m\AA$} & \colhead{} & \colhead{} & \colhead{$\AA$} &
\colhead{$km~s^{-1}$} } 
\startdata 
$9.391\pm0.002$ & $7.1\pm1.1$ & 44 & \mgxi & 9.169 & $123\pm64$ \\ 
$11.824\pm0.004$ & $5.5\pm3.3$ & 25 & \neix & 11.547 & $206\pm101$ \\ 
$12.430\pm0.001$ & $13.6\pm1.4$ & 180 & \nex & 12.134 & $80\pm24$ \\ 
$13.625\pm0.011$ & $5.0\pm1.8$ & 2 & $--$ & & \\
$13.774\pm0.001$ & $15.2\pm0.8$ & 168 & \neix & 13.447 & $107\pm22$ \\
$14.160\pm0.005$ & $9.1\pm 2.4$ & 22 & \fexvii & 13.827 & $167\pm106$ \\
$14.550\pm0.004$ & $11.1\pm1.6$ & 54 & \fexviii& 14.203 & $72\pm82$ \\
$15.000\pm0.005$ & $6.6\pm2.0$ & 10 & \fexvii & 15.015 & local \\
$15.375\pm0.003$ & $17.8\pm1.0$ & 10 & \fexvii & 15.015 & $199\pm59$ \\
$18.960\pm0.004$ & $9.4\pm4.1$ & 3 & \oviii & 18.969 & local\\
$16.396\pm0.002$ & $13.4\pm2.4$ & 70 & \oviii $\beta$ & 16.006 & $89\pm37$ \\ 
$19.080\pm0.005$ & $15.2\pm2.6$ & 31 & \ovii $\beta$ & 18.627 & $107\pm79$ \\ 
$19.430\pm0.004$ & $19.1\pm3.2$ & 122 & \oviii $\alpha$ & 18.969 & 
$110\pm62$ \\ 
$19.805\pm0.006$ & $14.7\pm2.5$ & 27 & $--$ & & \\ 
$19.845\pm0.005$ & $16.4\pm2.5$ & 34 & $--$ & & \\
$21.600\pm0.007$ & $12.2\pm1.9$ & 9 & \ovii & 21.602 & local \\
$22.128\pm0.008$ & $48.7\pm7.0$ & 49 & \ovii & 21.602 & $99\pm108$ \\
$22.560\pm0.005$ & $26.2\pm5.5$ & 24 & \ovi & 22.026 & $130\pm66$ \\
$20.250\pm0.005$ & $14.8\pm4.0$ & 20 & $--$ & & \\ 
$23.286\pm0.002$ & $19.1\pm4.7$ & 23 & \oii & 23.310 & local \\ 
$23.343\pm0.011$ & $25.4\pm5.8$ & 20 & $--$ & & \\ 
$23.505\pm0.001$ & $22.5\pm3.6$ & 40 & \oi & 23.448 & local \\ 
\enddata 
\tablenotetext{a} {Change in $\chi^{2}$
after inclusion of Gaussian over continuum model} \tablenotetext{b} {The
``$--$'' marked the features with no identification} \tablenotetext{c} 
{Atomic lines are from NIST database 
for all, except \oin-\oii are from \citet{Garcia2005} and \ovi is from 
\citet{Yao2009}}
\end{deluxetable}

\clearpage
\begin{deluxetable}{lcccc}
\tabletypesize{\scriptsize}
\tablewidth{0pt}
\tablecaption{Model parameters for the Ark~564 \chandra HETG and LETG spectra}
\tablehead{
\colhead{}           & \colhead{Units}      &
\colhead{MEG}          & \colhead{MEG+HEG} & \colhead{LETG}}
\startdata
\textbf{Powerlaw}\\
Photon Index ($\Gamma$)  &   & $2.34\pm0.04$ &      $2.38\pm0.06$  
& $2.85\pm0.05$    \\
Normalization &$10^{-3}~ph~keV^{-1}~s^{-1}~cm^{-2}$ & $13.4\pm0.3$ 
& $13.7\pm0.5$ & $20.8\pm0.9$ \\
\textbf{Black Body}\\
kT  & $keV$  &      $0.138\pm0.001$  &  $0.131\pm0.006$  &  $0.147\pm0.006$ \\
Normalization\tablenotemark{a} & $10^{-5}~L_{39}/D^{2}_{10}$  &   
 $47.7\pm1.6$       &      $49.3\pm2.4$    &  $32.5\pm2.3$  \\

\textbf{Warm Absorber: HIP}\\
$Log~U$     &             & $0.39\pm0.03$ & $0.39\pm0.04$   & $0.48\pm0.09$ \\
$Log~N_{H}$ & $cm^{-2}$   & $20.94\pm0.02$ & $20.89\pm0.03$   &  
$20.78\pm0.06$  \\
$V_{out}$  & $km~s^{-1}$  & $94\pm13$  & $78\pm15$   &   $83\pm26$  \\
$V_{turb}$ & $km~s^{-1}$  & $120\pm150$  & $160\pm150$   &  $87\pm150$  \\
$Log~T$\tablenotemark{b}  & K &    $5.20\pm0.01$ & $5.20\pm0.01$   &  
$5.23\pm0.07$  \\
$Log~T/U (\propto P)\tablenotemark{c}$ & K & $4.81\pm0.04$ & 
$4.81\pm0.04$   &   $4.75\pm0.19$  \\

\textbf{Warm Absorber: LIP}\\
$Log~U$     &             & $-0.99\pm0.13$  & $-1.04\pm0.22$   &  
$-1.33\pm0.09$  \\
$Log~N_{H}$ & $cm^{-2}$   & $20.11\pm0.06$ & $20.00\pm0.17$   &  
$20.51\pm0.02$   \\
$V_{out}$  & $km~s^{-1}$  &    $137\pm37$ &    $144\pm64$   &  $189\pm22$   \\
$V_{turb}$ & $km~s^{-1}$  &    $193\pm150$  &  $186\pm150$    &  $80\pm150$  \\
$Log~T$\tablenotemark{b}  &  K &    $4.49\pm0.02$  & $4.48\pm0.02$  
&  $4.40\pm0.09$  \\
$Log~T/U~(\propto P)\tablenotemark{c}$ &  & $5.48\pm0.14$ & 
$5.52\pm0.23$  &   $5.73\pm0.13$  \\
\textbf{$\chi^{2}$}& & $4674$ & $5131$   &  $1370$  \\
Degrees of freedom & & $3979$ & $6792$   &  $1184$  \\
\enddata
\tablenotetext{a}{where $L^{39}$ is the source luminosity 
in units of $10^{39}~erg~s^{-1}$ and $D_{10}$ is the 
distance to the source in units of 10 kpc.}
\tablenotetext{b}{Derived from the column 
density and ionization parameter, assuming photoionization equilibrium.}
\tablenotetext{c}{The pressure $P\propto n_{e} T$. Assuming that 
both phases lie at the same
distance from the central source $n_{e}\propto 1/U$, and $P\propto T/U$.}
\end{deluxetable}

\clearpage

\begin{deluxetable}{lccccccc}
\tabletypesize{\scriptsize}
\tablecaption{Ionic Column Densities Predicted By Models.}
\tablewidth{0pt}
\tablehead{
\colhead{} & \multicolumn{3}{c}{HETG} & \colhead{} &\multicolumn{3}{c}{LETG}\\
\cline{2-4} \cline{6-8}\\
\colhead{Ion} & \colhead{$logN_{HIP}$} & \colhead{$logN_{LIP}$} & 
\colhead{$logN_{SUM}$} & & \colhead{$logN_{HIP}$} & \colhead{$logN_{LIP}$} & 
\colhead{$logN_{SUM}$} 
}
\startdata

\caxiv & 14.86 & --    & 14.86 & &14.67 & -- &	14.67 \\
\sxiv &	 15.16 & --    & 15.16 & &15.15 & -- &	15.15\\
\caxv &	 14.69 & --    & 14.69 & &14.61 & -- &	14.61\\
\nvii &	 16.17 & 15.41 & 16.24 & &15.92 & 15.36 &	16.03\\
\oiv & 	 10.34 & 15.05 & 15.05 & &--    & 16.26 &	16.26\\
\ov & 	 12.46 & 15.97 & 15.97 & &11.96 & 16.80 &	16.80\\
\ovi & 	 14.09 & 16.33 & 16.33 & &13.68 & 16.88 &	16.88\\
\ovii &  16.44 & 16.75 & 16.92 & &16.12 & 16.88 &	16.95\\
\oviii & 17.37 & 15.80 & 17.38 & &17.14 & 15.53 &	17.15\\
\fexvii &  16.06 & -- & 16.06  & &15.90 & -- &	15.90\\
\fexviii & 15.86 & -- & 15.86  & &15.81 & -- &	15.81\\
\neix &  16.61 & 14.96 & 16.62 & &16.36 & 14.37 &	16.36\\
\nex & 	 16.68 & 13.18 & 16.68 & &16.54 & 12.19 &	16.54\\
\mgx & 	 15.59 & 13.63 & 15.49 & &15.21 & 12.87 &	15.21\\
\mgxi &  16.37 & 12.80 & 16.37 & &16.19 & 11.77 &	16.19\\
\mgxii &  15.74 & 10.34 & 15.74 & &15.67	& -- &	15.67\\
\sixi &  15.75 & 12.92  & 15.75 & &15.47	& 10.80 &	15.47\\
\sixii &  15.79 & 11.76 & 15.79	& &15.60 & -- &	15.60\\
\sixiii & 16.14 & --  & 16.14	& &16.05 & -- &	16.05\\
\enddata
\end{deluxetable}

\clearpage

\begin{figure}
\epsscale{.80}
\plotone{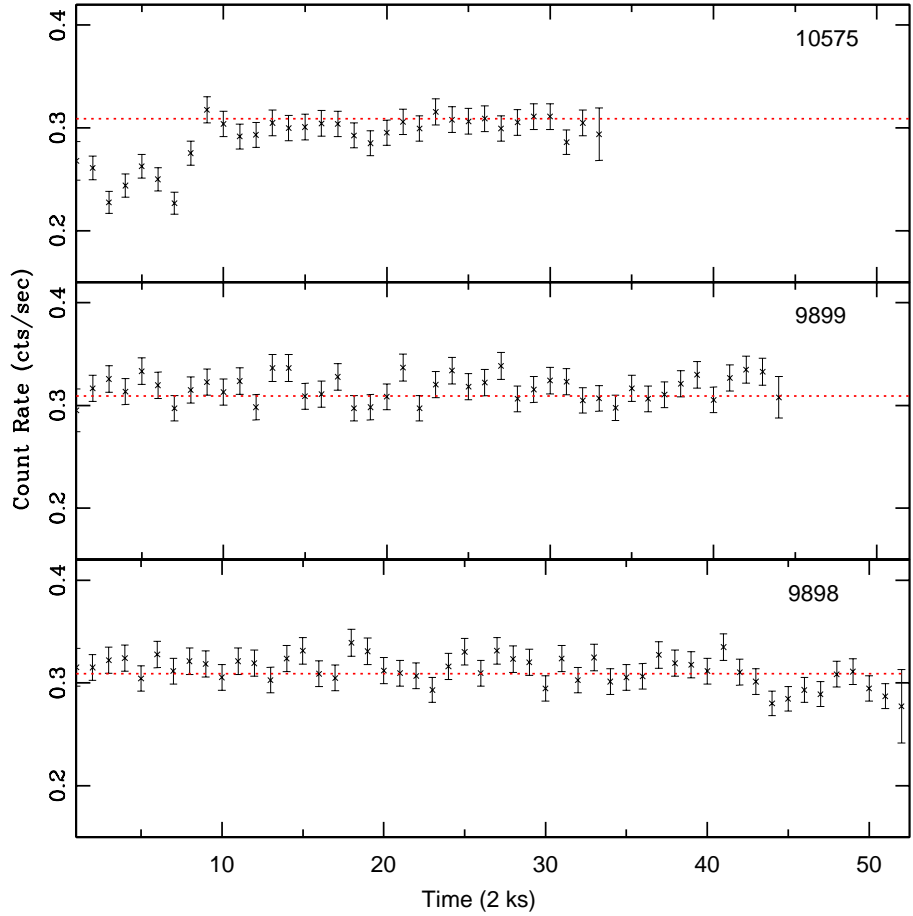}
\caption{Ark~564 light curve of the HETG observations 
analyzed in this work, binned at 2 ks resolution.  
Except for the initial 20 ks of obsID 10575, the count rates of 
all observations are consistent with each other, with average value 
of 0.31 cts s$^{-1}$ (dash curve). }
\end{figure}

\begin{figure}
\epsscale{.80}
\plotone{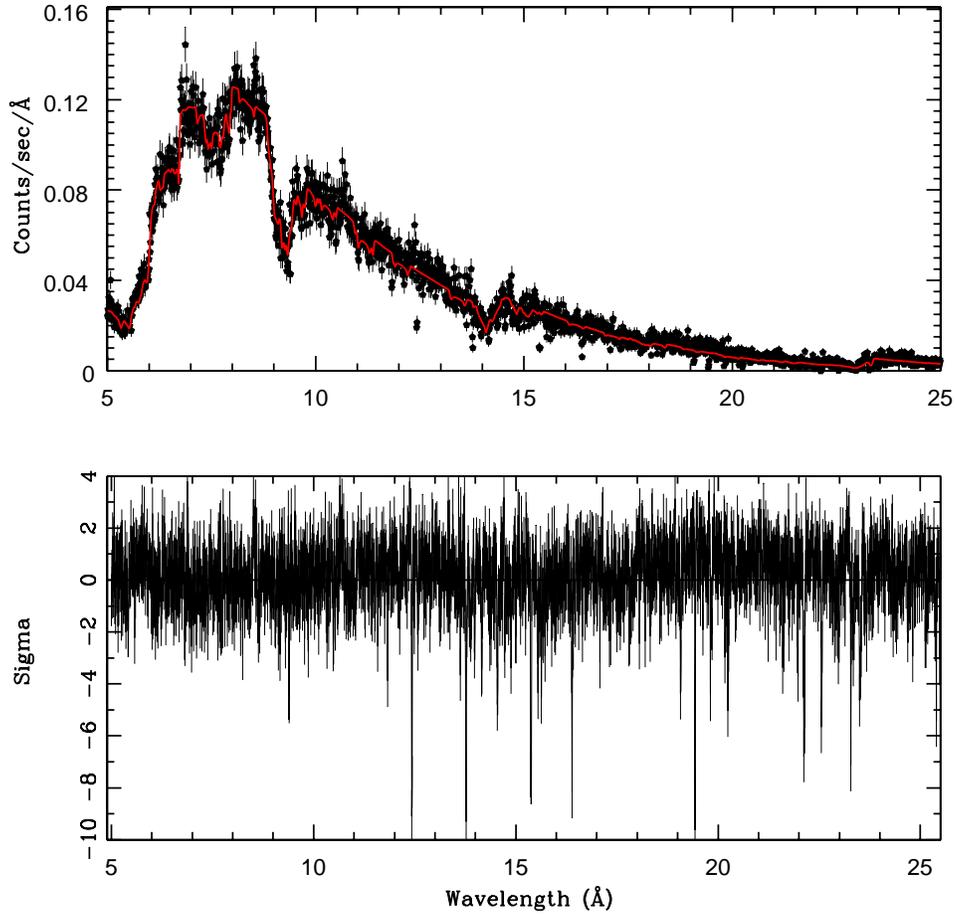}
\caption{{\it Top panel}: The Ark~564 co-added Medium Energy Grating
(MEG) spectrum in the observer frame. The red solid lines show the best
fit continuum model that consists of an absorbed power law and a black
body component.  {\it Bottom panel}: Plotted are the residuals showing
strong WAs features. }
\end{figure}

\begin{figure}
\epsscale{.80}
\plotone{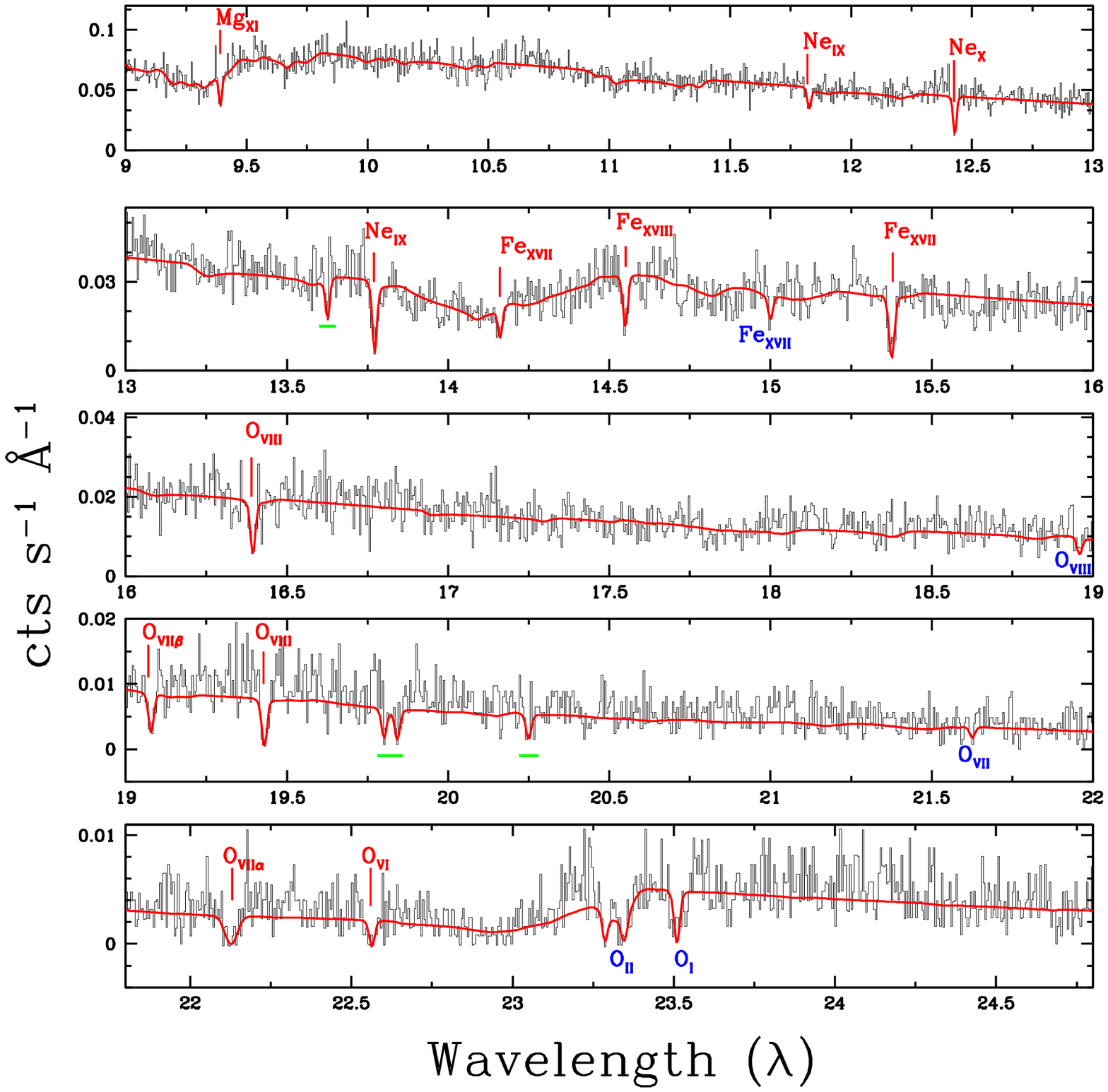}
\caption{Same as fig. 2; in addition to the continuum model, the
absorption features are fitted with Gaussians. Note the numerous warm
absorber features labeled in red, above the lines.  The local (z=0)
features are labeled in blue, below the lines. The unidentified features
are marked with green underlines.  The identification and interpretation
of these lines are discussed in a companion paper.}
\end{figure}

\begin{figure}
\epsscale{.80}
\plotone{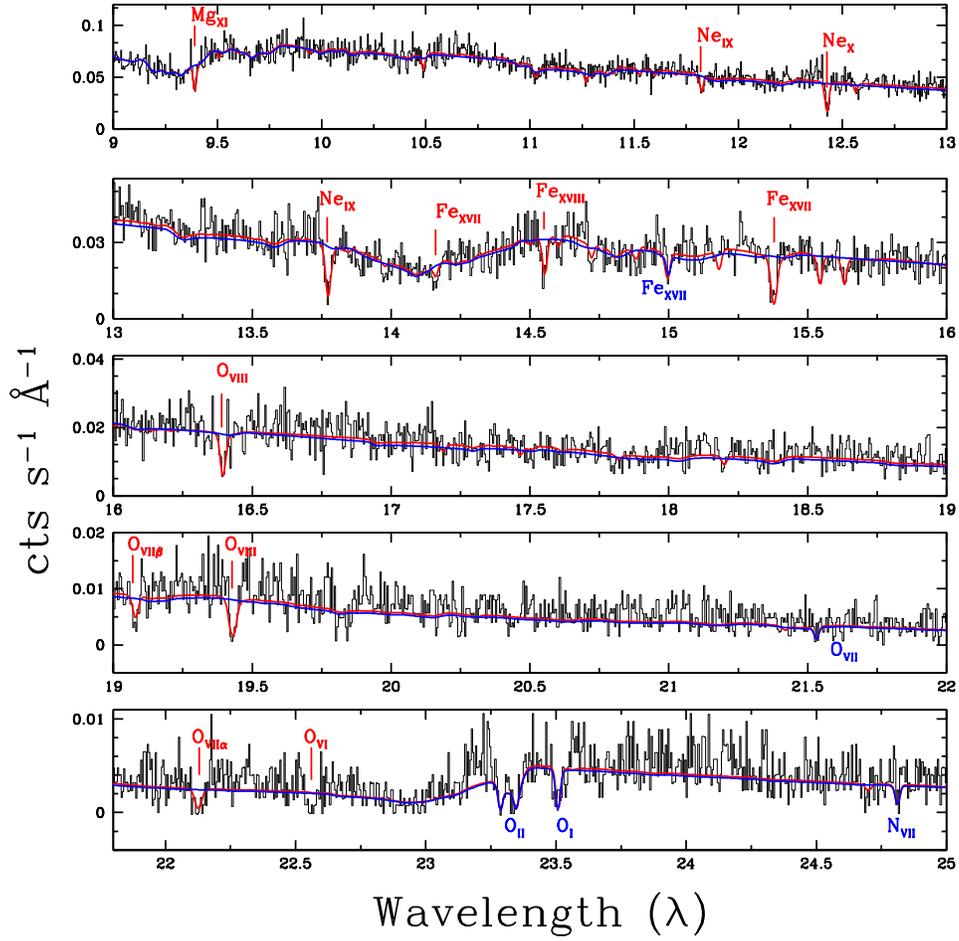}
\caption{Same as fig. 3, but the intrinsic absorption lines are modeled
with PHASE.  Only the high ionization phase (HIP) absorber is shown. }
\end{figure}

\begin{figure}
\epsscale{.80}
\plotone{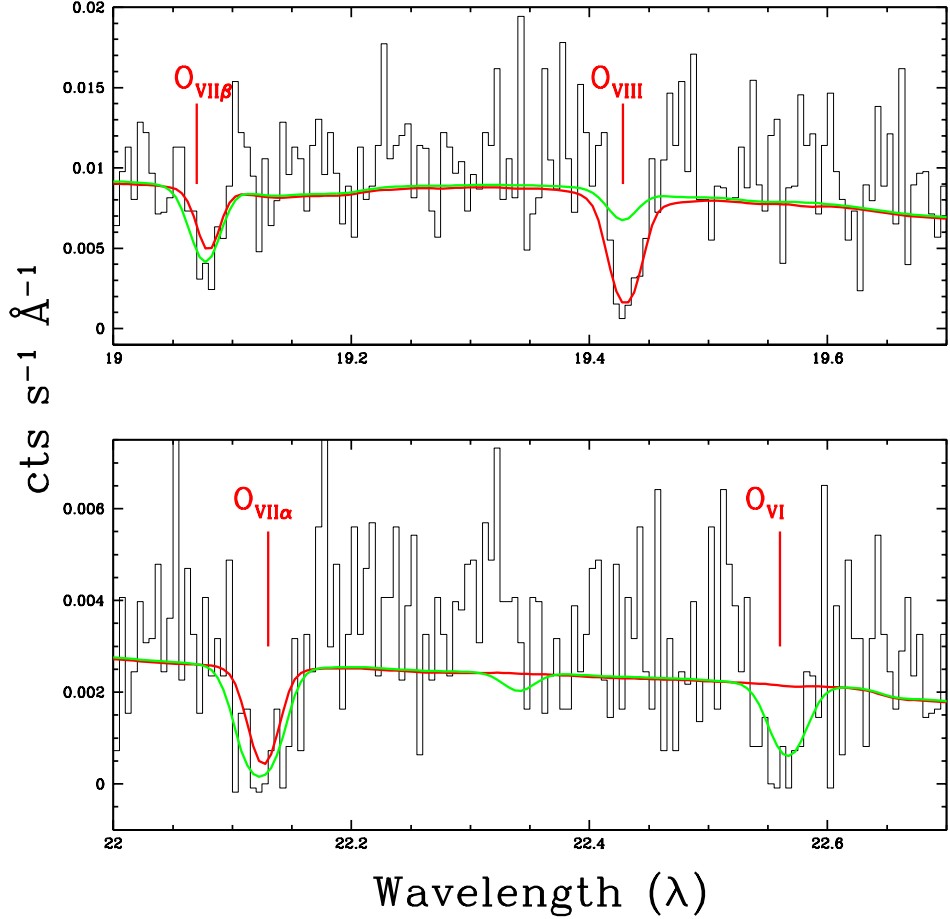}
\caption{An enlarged view of fig. 4 near the regions of \ovii $K\beta$ (top) 
and \ovii $K\alpha$ and \ovi (bottom). As can be observed, the \ovi line is 
not modeled by the HIP component (red solid curve) and it also underpredicts  
the \ovii absorption. The green dash curve shows the low ionization 
phase (LIP) of our model, reproducing the \ovi and \ovii absorption.}
\end{figure}

\begin{figure}
\epsscale{.80}
\plotone{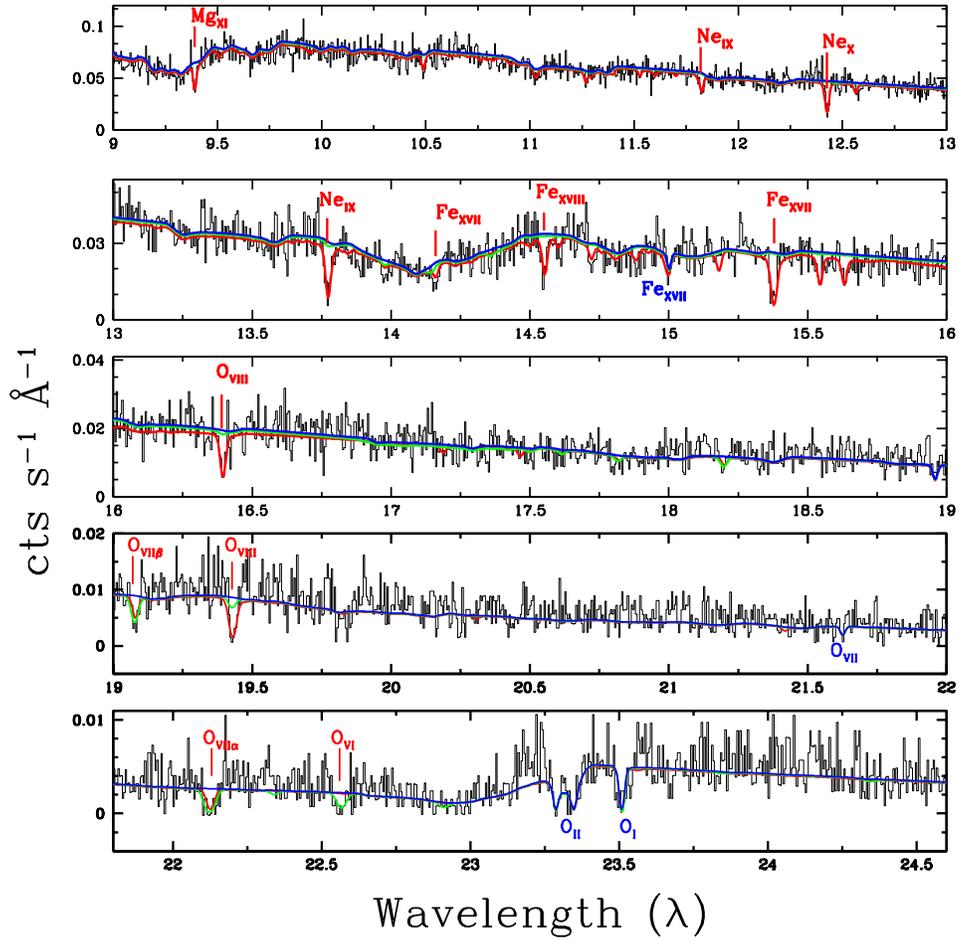}
\caption{Same as fig. 4, but showing both HIP (red) and LIP (green)
components of Model C. The LIP only contributes at lower energies,
particularly to \ovii and \ovi lines.}
\end{figure}

\begin{figure}
\epsscale{.80}
\plotone{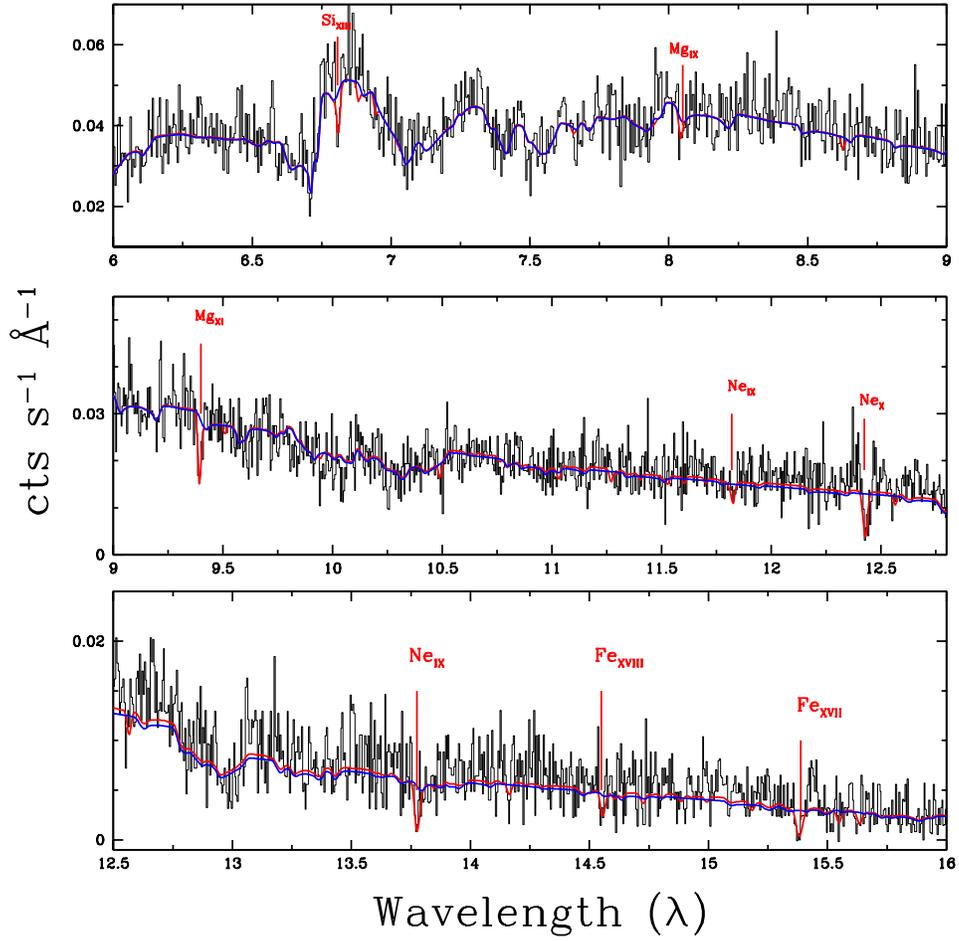}
\caption{The High Energy Grating (HEG) spectrum of 
Ark~564 in the observer frame. 
The blue and red lines show the continuum and the WA 
model respectively. All the intrinsic 
WA features are well modeled with the HIP absorber.}
\end{figure}

\begin{figure}
\epsscale{.80}
\plotone{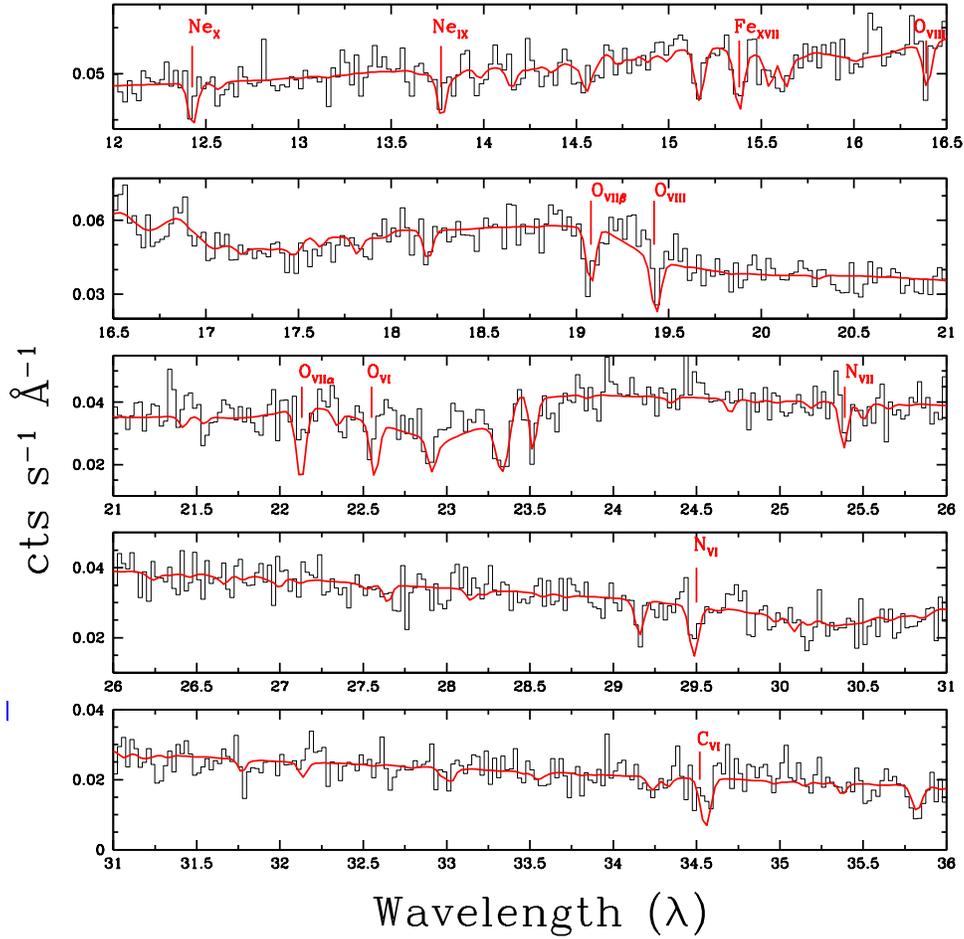}
\caption{The Low Energy Grating (LEG) spectrum of Ark~564 in the
observer frame, fitted with Model-C of the HETG. Though most of the WA
features are fitted well, but this model over estimates the \ovii
$K\alpha$ line at $22.13\AA$.}
\end{figure}

\begin{figure}
\epsscale{.80}
\plotone{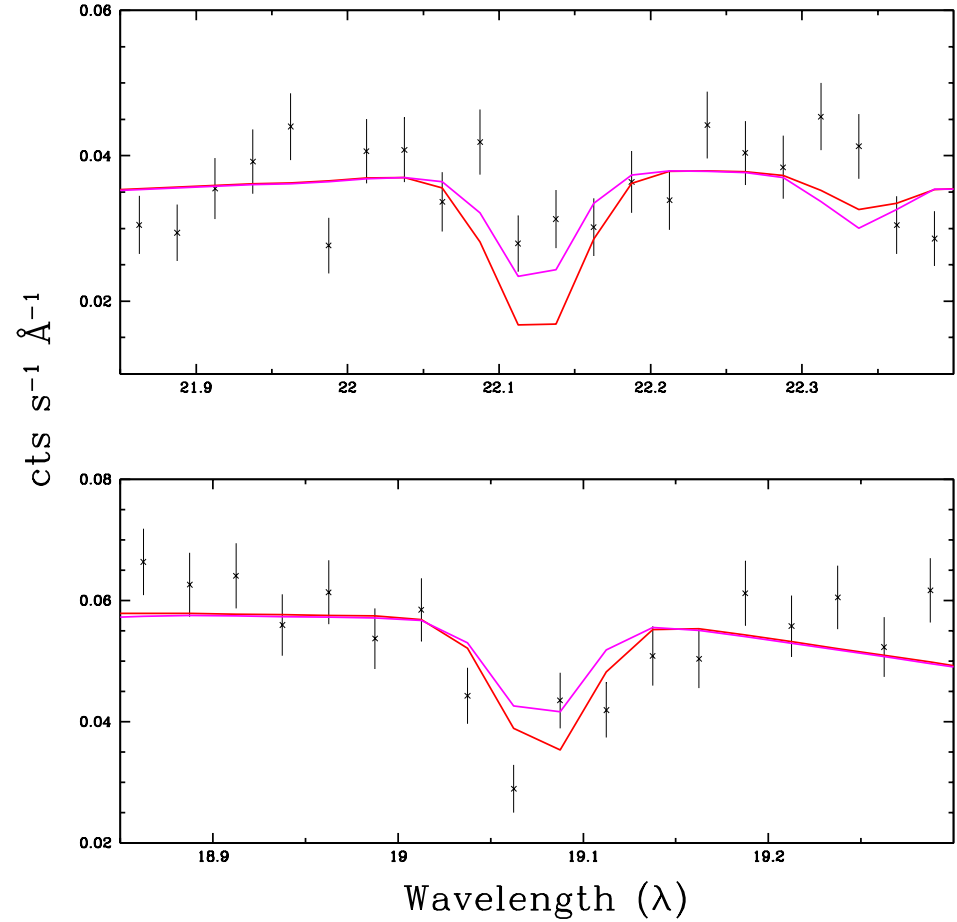}
\caption{An enlarged view of Fig. 8 near the region of \ovii $K\alpha$
(top) and \ovii $K\beta$ (bottom). The red (dotted line) and 
magenta (solid line) curve show the
best fit HETG and LETG models respectively. The HETG model overestimates
the \ovii k$\alpha$ (red curve, top panel), while the best fit LETG model
underestimates \ovii k$\beta$ (magenta curve, bottom panel)}
\end{figure}

\begin{figure}
\epsscale{.80}
\plotone{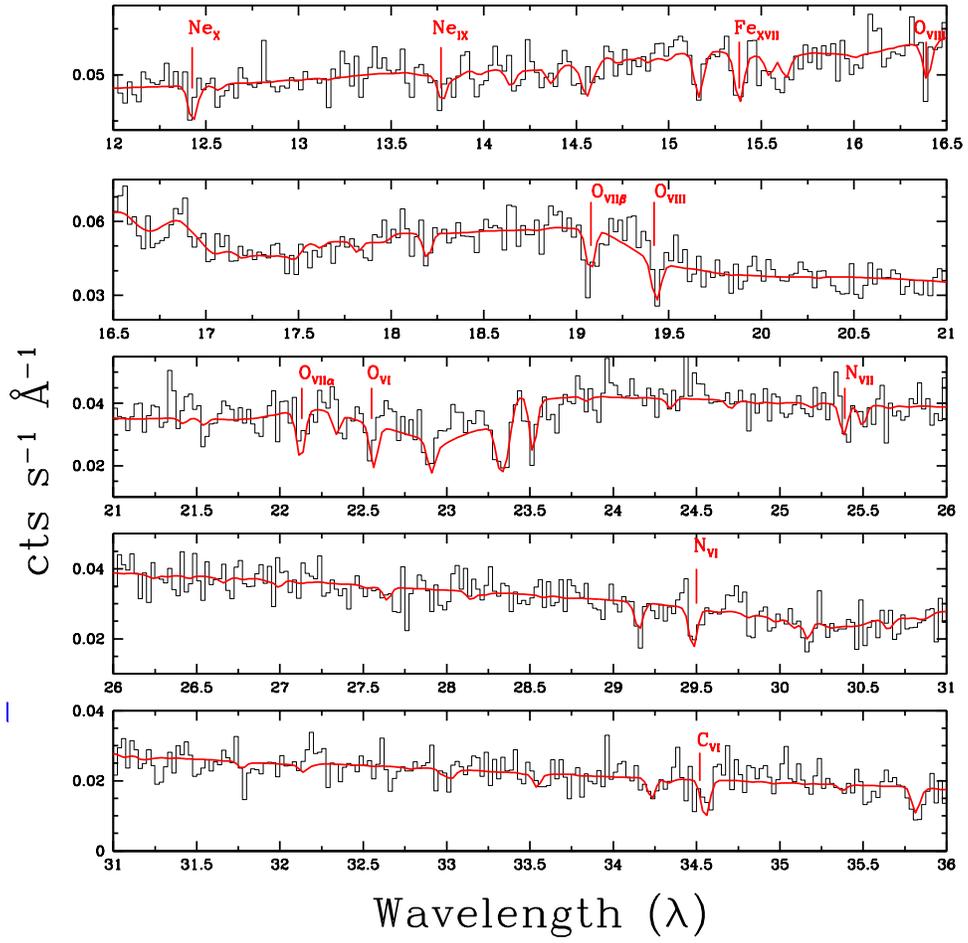}
\caption{Same as fig. 8, showing the LETG best fit two absorber model.}
\end{figure}

\begin{figure}
\epsscale{.80}
\plotone{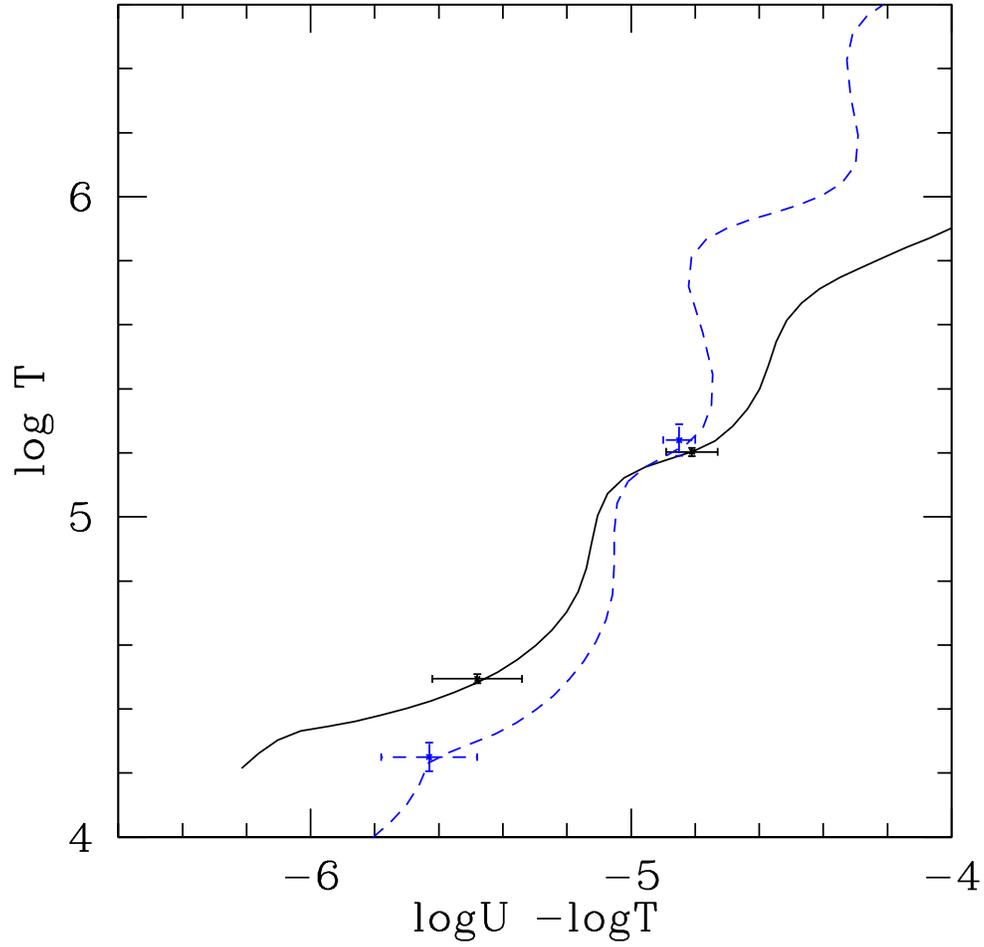}
\caption{Pressure equilibrium curve (S-curve) for the 
Ark~564 SED used in the present analysis. The black curve is for the solar 
metallicity while the dashed blue curve is for super-solar (10 solar) 
metallicity. The points are for the LIP (lower) and HIP (upper) components.}
\end{figure}

\end{document}